\documentclass{article}

\usepackage{verbatim}
\usepackage[T1]{fontenc}
\usepackage{amsfonts}
\usepackage{amsmath,amssymb,fancyhdr,amsthm,bm,cite,dsfont}
\usepackage{relsize}
\newtheorem{thm}{Theorem}[section]

\newtheorem{prop}[thm]{Proposition}

\theoremstyle{definition}

\theoremstyle{remark}

\def\lra{\leftrightarrow}
\def\ff{{\mathbf f}}

\def\muc{\overline{\mu}}

\def\pic{\overline{\pi}}

\def\rhoc{\overline{\rho}}

\def\s5{\sqrt{5}}

\def\u{{\boldsymbol u}}
\def\mm{{\boldsymbol m}}
\def\mmc{\overline{{\boldsymbol m}}}
\def\kk{{\mathbf k}}
\def\ll{{\boldsymbol l}}

\def\P{\Psi}

\def\bl {\mbox{\boldmath{$\ell$}}}
\def\bn {\mbox{\boldmath{$n$}}}

\def \bmm {\mbox{{$\bf m$}}}

\def\s{\sigma}

\def\zc{\overline{\zeta}}

\def\bar{\overline}

\newcommand{\rhob}{\bm{\rho}}
\newcommand{\Phib}{\bm{\Phi}}

\newcounter{mnotecount}[section]

\newcommand{\mnotex}[1]
{\protect{\stepcounter{mnotecount}}$^{\mbox{\footnotesize$
\bullet$\themnotecount}}$ \marginpar{
\raggedright\tiny\em
$\!\!\!\!\!\!\,\bullet$\themnotecount: #1} }

\def\tho{\textrm{\TH}}
\def\eth{\textrm{\dh}}

\newcommand{\be}{\begin{equation}}
\newcommand{\ee}{\end{equation}}
\newcommand{\bes}{\begin{equation*}}
\newcommand{\ees}{\end{equation*}}
\newcommand{\beqn}{\begin{eqnarray}}
\newcommand{\eeqn}{\end{eqnarray}}
\newcommand{\beq}{\begin{eqnarray*}}
\newcommand{\eeq}{\end{eqnarray*}}

\def\P{\Psi}

\def\frf{\mathfrak{f}}

\def\PP{\F^2}

\def\Phib{\boldsymbol{\Phi}}

\def\rhob{\boldsymbol{\rho}}

\def\PP{p}

\def\cf{\mathfrak{c}}
\def\df{\mathfrak{d}}
\def\ef{\mathfrak{e}}
\def\ff{\mathfrak{f}}

\def\bi#1#2#3#4#5{bi(#1#2#3#4#5)}

\def\psic{\overline{\psi}}
\def\zc{\overline{\zeta}}
\def\zetac{\overline{\zeta}}

\def\ups{\upsilon}
\def\upsc{\overline{\ups}}
\def\pic{\overline{\pi}}

\def\varsic{\overline{\varsigma}}

\def\tho{\textrm{\TH}}
\def\eth{\textrm{\dh}}

\def\Rfdrie{\Psi^*_3}
\def\Lfdrie{^*\Psi_3}
\def\Lfdriec{\overline{\Lfdrie}}
\def\Rfdriec{\overline{\Rfdrie}}

\def\bia#1{Bia#1}
\def\biac#1{$\overline{\textrm{Bia#1}}$}

\def\ric#1{Ric#1}

\def\Dh{\widehat{{\mathcal D}}}
\def\W{w}
\def\w{w}
\def\rone{\textrm{rank 1}}
\def\rdrie{\textrm{rank 3}}

\begin{document}

\title{Finalizing the classification of type II or more special Einstein  spacetimes in five dimensions}

\author{Lode Wylleman
\thanks{E-mail address: lwyllema@cage.ugent.be}
\\
Ghent University, Department of Mathematical Analysis\\
EA16, Galglaan 2, 9000 Ghent, Belgium
}
\maketitle

\begin{abstract} Einstein spacetimes in 5d that are of genuine type II in the null alignment classification are considered. It is shown that the unique geodesic multiple Weyl aligned null direction (mWAND) cannot have an optical matrix of rank 1 or 3. This finalizes the classification of all type II or more special Einstein spacetimes in 5d (i.e., those allowing for an mWAND), and also proves that they all satisfy the so-called optical constraint. 
\end{abstract}

\section{Introduction}
In the null alignment classification of the Weyl tensor by Coley, Milson, Pravda and Pravdov\'{a}~\cite{CMPP,OrtPraPra12rev}, an Einstein spacetime is type II or more special iff it allows for a multiple Weyl aligned null direction (mWAND) at each point.
We deal with Einstein spacetimes in five dimensions (5d) of this kind and report on a result that finalizes their classification. 

In \cite{PRAVDA-BIANCHI,DURKEE-GOLDBERG-SACHS,5DGS} the relevant part of the celebrated Goldberg-Sachs (GS) theorem~\cite{SKMHH} was suitably generalized for such spacetimes. First, there is always a geodesic multiple Weyl aligned null direction (mWAND) $\bl$~\cite{PRAVDA-BIANCHI,DURKEE-GOLDBERG-SACHS}. Moreover, all spacetimes that allow for a non-geodesic mWAND are of type D or O (the last case giving the spacetimes of constant curvature) and were fully obtained in \cite{DURKEE-GOLDBERG-SACHS}; we refer to their set as the Durkee-Reall class.  Second, define the optical matrix $\rhob$ of $\bl$ with respect to a real null frame $(\bmm_0=\bl,\bmm_1=\bn,\bmm_i)$ by $\rho_{ij}\equiv \nabla_j\ell_i$ ($i,j$ run from 2 to 4). By geodesy of $\bl$ the rank of $\rhob$ doesn't depend on the other frame vectors. In \cite{PRAVDA-BIANCHI} it was proven for the type N and III cases\footnote{A mentioned null alignment type always refers to a genuine type, i.e., not more special.~\cite{OrtPraPra12rev}.} that the unique mWAND has a rank 2 optical matrix which by rotating the $\bmm_i$ can be put in the form $\rhob^2$ below. For types II and D the matrix $\Phib$ with components $\Phi_{ij}=C_{0i1j}$ determines the boost weight 0 part of the Weyl tensor completely\cite{PRAVDA-TYPED}. In \cite{5DGS,DURKEE-TYPEII} it was shown that $\rhob$ and $\Phib$ can be simultaneously put in one of the following forms:
\beqn
\label{rhob1}
&\rhob^1=\begin{bmatrix}
0&0&0\\0&0&0\\\rho_{42}&\rho_{43}&\rho_{44}
\end{bmatrix},\qquad
&\Phib^1=\,\PP\,\begin{bmatrix}
-1&0&0\\
0&-1&0\\
0&0&1
\end{bmatrix};\\
\label{rhob2}
&\rhob^2=b\begin{bmatrix}
1&a&0\\
-a&1&0\\
0&0&0
\end{bmatrix},
\qquad
&\Phib^2=\begin{bmatrix}
\Phi_{22}&\Phi_{23}&0\\
\Phi_{32}&\Phi_{33}&0\\
\Phi_{42}&\Phi_{43}&0
\end{bmatrix};\\
\label{rhob3}
&\rhob^3=b\begin{bmatrix}
1&a&0\\
-a&1&0\\
0&0&1+a^2
\end{bmatrix},
\qquad
&\Phib^3=\PP\,\begin{bmatrix}
\cos(\vartheta)&\sin(\vartheta)&0\\
-\sin(\vartheta)&\cos(\vartheta)&0\\
0&0&1
\end{bmatrix},
\eeqn
where in the last case also
\beqn\label{rhob3-extra}
\vartheta\in[0,\pi[,\qquad \,a=\tan(\vartheta/2)\;\;\lra\;\;\cos(\vartheta)=\frac{1-a^2}{1+a^2},\,\;\sin(\vartheta)=\frac{2a}{1+a^2},
\eeqn
where the superscript $i$ indicates the rank of $\rhob\neq 0$. 

In \cite{DeFGodRea15} the spacetimes that allow for a geodesic mWAND with a rank 3 optical matrix were found to be of type D; we shall make a straightforward verification of this result in the present paper, not relying on coordinates as in \cite{DeFGodRea15}. The corresponding type D metrics have been exhaustively integrated, independently in \cite{DeFGodRea15} and \cite{WylPar15}.

The spacetimes allowing for a non-twisting geodesic mWAND with a rank 2 optical matrix were all obtained in \cite{ReaGraTur13} (see also \cite{PraPra08}). Regarding the remaining case of a twisting mWAND, the spacetimes of type D have been shown to be 1+4 Brinkmann warps in \cite{WylPar15}. The remaining cases where the mWAND is unique (types II, III and N) and twisting are dealt with in \cite{DeFGodRea15b}.

In the present paper we will show: 

\begin{prop}\label{prop:rank 1-3}
If a  type II or D Einstein spacetime in 5d allows for a geodesic mWAND with a rank 1 or 3 optical matrix then it is of type D.
\end{prop}

In the rank 1 case the spacetime thus belongs to the Durkee-Reall class. By the results of \cite{PRAVDA-BIANCHI} we get a first corollary:
\begin{thm}\label{thm:rank 1-3}
The unique (thus geodesic) multiple WAND of a genuine type II, III or N Einstein spacetime in 5d has an optical matrix of rank 2 or 0.
\end{thm}
Merging with the results of \cite{5DGS} we obtain a second corollary: 
\begin{thm}\label{thm:rank 1-3}
In 5d an Einstein spacetime of type II or more special always admits a multiple WAND that satisfies the optical constraint, i.e., for which $\rhob\rhob^t\propto\rhob+\rhob^t$.
\end{thm} 

\section{Proof of proposition \ref{prop:rank 1-3}}
\label{sec:proof-prop1-3}

Assume that a type II or D Einstein spacetime in 5d allows for a geodesic mWAND $\kk$ with an optical matrix of rank 1 or 3. We will respectively refer to these cases as the rank 1 and the rank 3 case whenever we need to distinguish between them, although a substantial part of the analysis runs in parallel.

For convenience we work in the 2+2+1 covariant, scalar, semi-complex GHP formalism introduced in \cite{WylPar15}. 
The vector field $\kk$ is geodesic,
\begin{align}
&\kappa=\df=0,\label{eq:k-geod}
\end{align}
and aligned with an mWAND at each point, such that the b.w.\ $>0$ components vanish:
\begin{align}
&\Psi_0=\Psi_1=0,\qquad {}^*\Psi_0={}^*\Psi_1=\Psi_1^*=0,\qquad \Psi_{00}=\Psi_{01}=0.\label{eq:bw-1-2zero}
\end{align}
We need 
In a type D spacetime we align $\bn$ with a second double WAND, such that also the b.w.\ $<0$ components vanish:
\begin{align}
&\label{eq:bw-1vanish}
{}^*\Psi_3=0,\qquad \Psi_3=0,\qquad \Psi_{12}=0,\qquad \Psi_3^*=0,\\
&\label{eq:bw-2cond}
\Psi_4=0,\qquad \Psi_{22}=0,\qquad \Psi^*_4=0.
\end{align}
A comparison between the notation used in this formalism and the one in \cite{PRAVDA-GHP}, for this situation, is given in table \ref{Table: link weylconn}.
The normalized forms (\ref{rhob1}) and (\ref{rhob3},\,\ref{rhob3-extra}) translate to the joint conditions (cf.\ Table \ref{Table: link weylconn}):
\begin{align}
&^*\Psi_{2}=\Psi_2^*=\Psi_{02}=0,\\
&\Psi_2=2\W\Psi_{11},\quad \overline{\W}=\frac{1}{\W}\qquad [\w=-e^{i\vartheta}]\,,\label{eq:Psi2para}\\
&\sigma=\eta=0\,,\\
&\rho=-\frac 12(\w-1)\ff\,,\label{eq:rhoeq}\\
\intertext{
and the separate conditions}
& \text{rank}(\rhob)=1: \qquad \w=1,\qquad \rho=0,\qquad (\varsigma,\ff)\neq (0,0);\label{rank1cond}\\
& \text{rank}(\rhob)=3: \qquad \w\neq 1,\qquad \varsigma=0,\qquad \ff\neq 0.\label{rank3cond}
\end{align}

\begin{table}[ht]
\begin{tabular}{llllll}
&b.w.\ $0$ Weyl comps&&b.w.\ $-1$ Weyl comps&&b.w.\ $-2$ Weyl comps\\
\hline\\
$\Psi_2$&$\frac 12({\Phi_{22}+\Phi_{33}})+i\Phi_{[23]}$&$\Psi_3$&$\frac{1}{\sqrt{2}}(\Psi'_{323}-i\Psi'_{223})+\frac{1}{2\sqrt{2}}(\Psi'_{424}+i\Psi'_{434})$
&$\Psi_4$&$\frac{1}{2}(\Psi'_{22}-\Psi'_{33})+i\Psi'_{(23)}$\\
$\Psi_{11}$&$-\frac 12{\Phi_{44}}$&&&$\Psi_{22}$&$\frac12\Psi'_{44}$\\
${}^*\Psi_2$&$-\frac{1}{\sqrt{2}}(\Phi_{42}+i\Phi_{43})$&${}^*\Psi_3$&$-\frac{1}{2}(\Psi'_{224}-\Psi'_{334})-i\Psi'_{(23)4}$&&\\
$\Psi^*_2$&$-\frac{1}{\sqrt{2}}(\Phi_{24}-i\Phi_{34})$&$\Psi^*_3$&$-\frac{1}{2}(\Psi'_{224}+\Psi'_{334})+i\Psi'_{[23]4}$
&$\Psi^*_4$&$\frac{1}{\sqrt{2}}(\Psi'_{24}+i\Psi'_{34})$\\
$\Psi_{02}$&$-\frac12(\Phi_{22}-\Phi_{33})+i\Phi_{(23)}$&$\Psi_{12}$&$\frac{1}{2\sqrt{2}}(\Psi'_{424}-i\Psi'_{434})$&&\\
\end{tabular}\\
\vspace{.3cm}\\
\begin{tabular}{llllll}
&\quad $\widetilde{\Gamma}_{a1b}$&&$\widetilde{\Gamma}_{a2b}$&&$\widetilde{\Gamma}_{35b}$\\
\hline
$\kappa$&$-\frac{1}{\sqrt{2}}(\kappa_{2}-i\kappa_{3})$&$\nu$&$\frac{1}{\sqrt{2}}(\kappa'_{2}+i\kappa'_{3})$
&$\chi$&$\frac{1}{\sqrt{2}}(\stackrel{4}{M}_{20}-i\stackrel{4}{M}_{30})$\\
$\frak d$&$-\kappa_4$ &$\frak b$&$\kappa'_4$&&\\
$\tau$&$\frac{1}{\sqrt{2}}(\tau_{2}-i\tau_{3})$&$\pi$&$-\frac{1}{\sqrt{2}}(\tau'_{2}+i\tau'_{3})$
&$\omega$&$-\frac{1}{\sqrt{2}}(\stackrel{4}{M}_{21}-i\stackrel{4}{M}_{31})$\\
$\frak e$&$\tau_4$&$\frak a$&$-\tau'_4$&&\\
$\sigma$&$-\frac{1}{2}(\rho_{22}-\rho_{33})+i\rho_{(23)}$&$\lambda$&$-\frac{1}{2}(\rho'_{22}-\rho'_{33})-i\rho'_{(23)}$
&$\phi$&$\frac{1}{{2}}(\stackrel{4}{M}_{22}-\stackrel{4}{M}_{33})-i\stackrel{4}{M}_{(23)}$\\
$\rho$&$-\frac{1}{2}(\rho_{22}+\rho_{33})-i\rho_{[23]}$&$\mu$&$-\frac{1}{2}(\rho'_{22}+\rho'_{33})+i\rho'_{[23]}$
&$\upsilon$&$\frac{1}{{2}}(\stackrel{4}{M}_{22}+\stackrel{4}{M}_{33})+i\stackrel{4}{M}_{[23]}$\\
$\eta$&$-\frac{1}{\sqrt{2}}(\rho_{24}-i\rho_{34})$&$\zeta$&$-\frac{1}{\sqrt{2}}(\rho'_{24}+i\rho'_{34})$
&$\psi$&$\frac{1}{\sqrt{2}}(\stackrel{4}{M}_{24}-i\stackrel{4}{M}_{34})$\\
$\varsigma$&$-\frac{1}{\sqrt{2}}(\rho_{42}-i\rho_{43})$&$\xi$&$-\frac{1}{\sqrt{2}}(\rho'_{42}-i\rho'_{43})$&&\\
$\frak f$&$-\rho_{44}$&$\frak c$&$-\rho'_{44}$&&\\
\end{tabular}\\
\caption{The b.w.\ $\leq0$ Weyl components and the connection coefficients used in the 2+2+1 GHP formalism in terms of the symbols used in \cite{PRAVDA-GHP}, given the relation \eqref{link snp nullframe}.}\label{Table: link weylconn}
\end{table}


\noindent {\em Principle of proof.} We need to show that there is a second mWAND. Let me point out the principle of the proof and anticipate on the result. Any null direction different from $\kk$ can be obtained by null rotating $\ll$ about $\kk$. Such a null rotation is defined by its action
\be
\begin{aligned}
&\kk\mapsto \kk,\quad \ll\mapsto\ll+B\mm +\bar{B}\mmc
+C \u+\left(B\bar{B} + \tfrac 12 C^{2}\right)\kk,\\
&\mm\mapsto\mm+\bar{B}\kk,\quad \mmc\mapsto\mmc+B
\kk,\quad \u\mapsto \u+C\kk
\end{aligned} \label{eq:null-rotation}
\ee
on the original frame, where $B$ is a complex and $C$ a real null rotation parameter. Under \eqref{eq:null-rotation} the b.w. $\geq 0$ Weyl components are invariant~\cite{ALIGNMENT}, while the b.w. $<0$ components change according to
\begin{align}
&\Psi_3\mapsto\Psi_3+6B\W\Psi_{11},\quad
\Psi_{12}\mapsto\Psi_{12}+2\bar{B} \Psi_{11},\quad
{}^*\Psi_3\mapsto{}^*\Psi_3,\quad
\Psi_3^*\mapsto\Psi_3^* -2C(\W-1)\Psi _{11};\label{bw-1compchange}\\
&\Psi_4\mapsto\Psi_4+12 B^{2}\W\Psi_{11},\quad \Psi_{22}\mapsto\Psi_{22}+\left(4 B\bar{B}+ C^{2}\frac{(\W-1)^2}{\W}\right)\Psi_{11},\quad\Psi _4^*\mapsto\Psi _4^*-6 B C (\W-1)\Psi_{11}.\label{bw-2compchange}
\end{align}
The new direction $\ll$ is a second double WAND if and only if all transformed components of b.w.\ $<0$ vanish. From \eqref{bw-1compchange} we see that the b.w. $-1$ combinations ${}^*\Psi_3$, $\Psi_3-3\W\overline{\Psi_{12}}$, and $\Psi_3^*$ ($\W=1$) or $\Psi_3^*+\W\overline{\Psi_3^*}$ ($\W\neq 1$) are invariant under all null rotations \eqref{eq:null-rotation}. Hence there can only be a second mWAND if
\be\label{eq:bw-1cond}
{}^*\Psi_3=0,\qquad \Psi_3-3\W\overline{\Psi_{12}}=0, \qquad \Psi_3^*=0\;\; (\W=1) \quad\textrm{or}\quad \Psi_3^*+\W\overline{\Psi_3^*}=0\;\; (\W\neq 1).
\ee
If these conditions are met, we see from \eqref{bw-1compchange} that in the rank 3 case ($\W\neq 1$) the null rotation \eqref{eq:null-rotation} given by~\footnote{Notice that the expression for $C$ is real by the last condition in \eqref{eq:bw-1cond}, as required.}
\be\label{BC-paras}
B=-\frac{\Psi_3}{6\W\Psi_{11}}, \qquad C=\frac{\Psi_3^*}{2(\W-1)\Psi_{11}}
\ee
transforms to a unique frame in which all b.w.\ $-1$ Weyl components are zero, i.e.\ \eqref{eq:bw-1vanish} holds;
in the rank 1 case, \eqref{eq:bw-1vanish} is realized by a one-parameter class of frames, corresponding to the same value of $B$ but arbitrary $C$. Hence, in the rank 3 case there can only be one other mWAND, which is the case if and only if also the b.w.\ $-2$ components vanish in the unique frame, i.e.\ \eqref{eq:bw-2cond} holds;
in the rank 1 case there will be another mWAND (and then an infinite number of mWANDs, obtained by residual null rotations \eqref{eq:null-rotation} with $B=0$ and arbitrary $C$) if and only if \eqref{eq:bw-2cond} holds in any frame from the one-parameter class. The outcome for the rank 1 case is consistent with \cite[Theorem 2]{DURKEE-GOLDBERG-SACHS}: the resulting type D spacetimes are foliated by 3d Lorentzian submanifolds, and any null direction tangent to these manifolds is a mWAND.

To prove the type D condition it is thus necessary and sufficient to show that condition \eqref{eq:bw-1cond} is met, and that \eqref{eq:bw-2cond} holds in a frame that realizes \eqref{eq:bw-1vanish}.

Consider the rank 3 subcase $\W=-1$, corresponding to $\vartheta=0$ in (\ref{rhob3},\,\ref{rhob3-extra}). Then $\rho_{ij}\propto\delta_{ij}$,
i.e., $\kk$ is non-rotating and shearfree. Hence the corresponding Einstein spacetimes are Robinson-Trautman solutions, which are of type D (in dimension 5 these reduce to generalized Schwarzschild solutions, allowing for a possible cosmological constant and with planar, spherical or hyperbolic symmetry)~\cite{PodOrt06}. Hence the proposition is true for this case, so henceforth we assume
\begin{equation}\label{eq:wnot-1}
\w\neq -1.
\end{equation}

Regarding the rank 1 case, the type D condition has been shown to hold in the non-rotating subcase $\Sigma=0$, albeit in a tedious way relying on intricate coordinate descriptions.
Hence we could assume $\Sigma\neq 0$ here, but instead will stick to
the weaker assumption $(\Sigma,\ff)\neq (0,0)$ and prove the type D condition for rank 1 in a direct way, both for the rotating and non-rotating subcases and not relying on coordinate constructions.

\noindent{Proof of the theorem.} We split the proof into two parts. In part A we use Bianchi equations of b.w.\ 1 or 0 (together with appropriate Ricci equations) to show that the first two conditions in \eqref{eq:bw-1cond} hold, i.e.,
\be\label{eq:firstpart}
{}^*\Psi_3=0,\qquad \Psi_3-3\W\overline{\Psi_{12}}=0,
\ee
and then proceed by asserting the last condition in \eqref{eq:bw-1cond} for both the rank 1 and 3 cases. As explained above we can then null rotate to a frame in which all b.w.\ -1 components vanish. 
In part B we then show, using Bianchi equations of b.w.\ -1, that the b.w.\ -2 components vanish as well; again we do this in two stages: first $\Psi_{22}=\Psi_4^*=0$ and finally $\Psi_4=0$.\\


\noindent {\bf A. The conditions \eqref{eq:bw-1cond} hold.}\\ 

\noindent {\em A.1. \eqref{eq:firstpart} holds.} 
\begin{proof} We obtain the result by using appropriate Ricci and b.w.\ 1 or 0 Bianchi equations. To start with
\bia{7} reads $2\chi\Psi_{11}(\W+1)=0$, such that
$$
\chi=0
$$
follows from our assumption \eqref{eq:wnot-1}.
We obtain an algebraic constraint from \bia{15}:
\beqn
\label{eq:phi}
&&2(\W+1)\phi\P_{11}+\frak{f}\,\Lfdriec-2\varsigma \P_{12}=0.
\eeqn
\bia{16}, and \bia{17} and \bia{18}, respectively produce an expression for $\eth\Psi_2$ and two expressions for $2\eth\Psi_{11}$:
\begin{align}
&\eth\Psi_2=(3\Psi_2+2\Psi_{11})\tau+\Lfdriec\varsic-\Psi^*_3\varsigma-2\rho\Psi_{12},\label{d3Psi2}\\
&2\eth\Psi_{11}= (\ff-2\rhoc)\Psi_{12}+(\Rfdrie+\Rfdriec)\varsigma -\ff \overline{\Psi_3}+\psi\overline{\Psi_2}+(2\psi+4\tau)\Psi_{11},\label{d3Psi11-1}\\
&2\eth\Psi_{11}= (-\ff-2\rhoc+4\rho)\Psi_{12}+(\Rfdrie+\Rfdriec)\varsigma -\ff \overline{\Psi_3}+\psi(\overline{\Psi_2}+\Psi_2)+4\psi\Psi_{11}-2\varsic\Lfdriec.\label{d3Psi11-2}
\end{align}
Comparing \eqref{d3Psi11-1} with \eqref{d3Psi11-2} and using (\ref{eq:Psi2para},\,\ref{eq:rhoeq}) leads to a second algebraic constraint
\beqn
\label{eq:psitau}
&&[(\W+1)\psi-2\tau]\Psi_{11}-\varsic\,\Lfdriec+\W\frak{f}\Psi_{12}=0.
\eeqn
\ric{1}, \ric{22}, \ric{23}, \ric{36}, \ric{38}, and \bia{11}, \bia{13}, \biac{26}, \biac{27}, in combination with (\ref{eq:phi},\,\ref{d3Psi11-1}), yield
\begin{eqnarray}
&&\tho\varsigma=\frac{3\W-1}{2\W}\frak{f}\varsigma,\quad
\tho\varsic=-\frac12(\W-3)\frak{f}\varsic,\quad \tho \frak{f}=\frak{f}^2,\label{eq:thosigma}\\
&& \tho\tau=-\frac12(\W-1)(\pic+\tau),\quad \tho\psi=\frak{f}(\pic+\psi),\quad
\tho\phi=\varsigma(\pic+\psi)+\frac{\W-1}{2\W}\ff\phi,\\
&&\tho\Psi_{11}=-\ff\frac{(\W-1)^2}{\W}\Psi_{11},\qquad \tho\W=-\frac12\ff(\W^2-1),\label{eq:thoPsi11W}\\
&&\tho\Lfdriec=\frac{\W-1}{2\W}(\frak{f}\,\Lfdriec-4(\W+1)\phi\P_{11})-\varsigma(\overline{\Psi_3}-3\Psi_{12}),\\
&&\tho\Psi_{12}=\left(2\pic+2\tau-\frac{\W^2-1}{\W}\psi\right)\Psi_{11}-\frac 12
\frak{f}(\overline{\Psi_3}-3\P_{12}).\label{eq:thoPsi12}
\end{eqnarray}
The second equation of \eqref{eq:thoPsi11W} implies that {\em $\w$ cannot be constant in the rank 3 case}.
Acting with $\tho$ on the algebraic constraints (\ref{eq:phi},\,\ref{eq:psitau}),
substituting (\ref{eq:thosigma}-\ref{eq:thoPsi12}), and using
(\ref{eq:phi},\,\ref{eq:psitau})
to simplify the results,
one obtains
\beq
&&\rone:\quad (\ff^2+2\varsigma\varsic)\Lfdriec=0,\qquad (\ff^2+2\varsigma\varsic)(\Psi_3-3\overline{\Psi_{12}})=0;\\
&&\rdrie:\quad (3\W-1)\ff^2\Lfdriec=0,\qquad  \ff^2(\Psi_3-3\W\overline{\Psi_{12}}).
\eeq
Since $(\varsigma,\frf)\neq(0,0)$ and $\frf\neq 0,\w\neq\text{const}$ in the respective cases, we conclude that \eqref{eq:firstpart} holds.
\end{proof}

\noindent {\em A.2. The third condition in \eqref{eq:bw-1cond} holds.}
\begin{proof}
To show this we also invoke the zero-weighted Bianchi equations \bia{20},\,\bia{21},\,\bia{23}, and \biac{23}, which are
\begin{align}
&\Dh\Psi_2-\Dh\overline{\Psi_2}=[(2\Psi^*_3+\overline{\Psi_3^*})\rho
+(\Psi_3-3\overline{\Psi_{12}})\varsigma-\frf\Psi_3^*
-\ups(\Psi_2-\overline{\Psi_2})] \;-\;\textrm{c.c.}\label{eq:bi20sp}\\
&\Dh\Psi_2+\Dh\overline{\Psi_2}+2\Dh\Psi_{11}=[(2\Psi^*_3-\overline{\Psi_3^*})\rho+(\Psi_3+\overline{\Psi_{12}})\varsigma-\ups(\Psi_2+\overline{\Psi_2}+4\Psi_{11})]\;+\;\textrm{c.c.}\label{eq:bi21sp}\\
&\Dh\Psi_2=(2\rho-\frf)\Psi_3^*+(\Psi_2-2\Psi_{11})\ef-(\Psi_2+2\Psi_{11})\ups+2\varsic\Psi_{12},\label{eq:d5Psi2sp}\\
&\Dh\overline{\Psi_2}=(2\rhoc-\frf)\overline{\Psi_3^*}+(\overline{\Psi_2}-2\Psi_{11})\ef-(\overline{\Psi_2}+2\Psi_{11})\upsc+2\varsigma\overline{\Psi_{12}}.\label{eq:d5Psi2spc}
\end{align}
and also use \eqref{eq:firstpart} whenever needed.\\

\noindent {\em Rank 1 case.} Substitute $\Psi_2=2\Psi_{11}$ everywhere, and eliminate $\Dh\Psi_{11}$ and $\upsc$ from \eqref{eq:bi21sp}-\eqref{eq:d5Psi2spc}
; this leads to the algebraic equation
\be\label{eq:waq}
2(\varsic\Psi_{12}-2\ups\Psi_{11})+\frf(\Psi_3^*+2\overline{\Psi_3^*})=0.
\ee
Since $\rho=0$ we also get $\frf\ups+(\tau-\psi)\varsic=0$ from \ric{10}; eliminating $\tau-\psi$ by means of \eqref{eq:psitau} we obtain $\frf(\varsic\Psi_{12}-2\ups\Psi_{11})=0$, such that by comparing with  \eqref{eq:waq}:
\be\label{ffPsi3st}
\frf(\Psi_3^*+2\overline{\Psi_3^*})=0\quad\Rightarrow\quad \ff\Psi_3^*=0.
\ee
Second, \eqref{d3Psi2} also gives an expression for $2\eth\Psi_{11}$; putting it next to \eqref{d3Psi11-2}, where we use \eqref{eq:psitau} to eliminate $\psi$, gives
$$
2\eth\Psi_{11}=8\tau\Psi_{11}-\varsigma\Psi_3^*,\qquad 2\eth\Psi_{11}=8\tau\Psi_{11}+\varsigma(\Psi_3^*+\overline{\Psi_3^*}).
$$
Hence,
\be\label{varsigmaPsi3st}
\varsigma(\Psi_3^*+2\overline{\Psi_3^*})=0\quad\Rightarrow\quad \varsigma\Psi_3^*=0.
\ee
Since $(\varsigma,\frf)\neq (0,0)$ we arrive at $\Psi_3^*=0$, as we needed to show.\\

\noindent {\em Rank 3 case.} Here $\varsigma=0$, such that compatibility with \ric{26} requires
\be\label{eq:biri51432}
(\W-1)\ef+\W\,\upsc-\ups=0.
\ee
If we now substitute \eqref{eq:d5Psi2sp} and \eqref{eq:d5Psi2spc} into \eqref{eq:bi20sp}, and use \eqref{eq:biri51432} to simplify the result, we obtain
\be
\ff(\w-1)(\Psi_3^*+\w\overline{\Psi_3^*})=0
\ee
and are done immediately.
\end{proof}

We now null-rotate to a frame in which \eqref{eq:bw-1vanish} holds. Then \eqref{d3Psi2}-\eqref{eq:d5Psi2spc} is found to imply
\begin{align}
&\label{eq:taueq} \tau=\frac12(\W+1)\psi,\\
&\label{eq:ueq} \upsilon=\frac{\W-1}{\W+1}\ef,\\
&\label{eq:d3w} \eth\w=\frac{\psi}{2}(\w^2-1),
\end{align}

\noindent{\bf B. The conditions \eqref{eq:bw-2cond} hold.}\\

\noindent{\em B.2. $\Psi_{22}=\Psi_4^*=0$.}

\begin{proof} Here we shall use the $(-1,0)$-weighted Bianchi equations \bia{28},\biac{28}, \bia{29},\biac{29}, \bia{30} and the $(-1,1)$-weighted ones, viz.\ the complex conjugates of \bia{32}-\bia{35}. Given \eqref{eq:bw-1cond} the respective sets are
\begin{align}
&\tho\Psi_{22}-\tho'\Psi_2=\mu(3\Psi_2-2\Psi_{11})+\rhoc\Psi_{22}-\varsigma\Psi_4^*,\label{eq:bw-1sw0-a}\\
&\tho\Psi_{22}-\tho'\overline{\Psi_2}=\muc(3\overline{\Psi_2}-2\Psi_{11})+\rho\overline{\Psi_{22}}-\varsic\overline{\Psi_4^*},\label{eq:bw-1sw0-b}\\
&2\tho'\Psi_{11}=\cf(\overline{\Psi_2}-2\Psi_{11})+(2\rhoc-\frf)\Psi_{22}-4\mu\Psi_{11}-\varsigma\Psi_4^*,\label{eq:bw-1sw0-c}\\
&2\tho'\Psi_{11}=\cf({\Psi_2}-2\Psi_{11})+(2\rho-\frf)\Psi_{22}-4\muc\Psi_{11}-\varsic\overline{\Psi_4^*},\label{eq:bw-1sw0-d}\\
&2\tho\Psi_{22}-2\tho'\Psi_{11}=2\frf\Psi_{22}-\cf(\Psi_2+\overline{\Psi_2}-4\Psi_{11}),\label{eq:bw-1sw0-e}
\end{align}
and
\begin{align}
&\tho\overline{\Psi^*_4}=(\overline{\Psi_2}+2\Psi_{11})\omega-(3\overline{\Psi_2}-2\Psi_{11})\zetac+\overline{\Psi^*_4}\frf,\label{eq:bw-1sw1-a}\\
&\tho\overline{\Psi^*_4}=({\Psi_2}+2\Psi_{11})\omega-(\Psi_2+\overline{\Psi_2}-4\Psi_{11})\xi+\overline{\Psi^*_4}\rhoc+2\Psi_{22}\varsigma,\label{eq:bw-1sw1-b}\\
&0=(\Psi_2+\overline{\Psi_2}-4\Psi_{11})\omega-(3\overline{\Psi_2}+2\Psi_{11})\zetac+(\overline{\Psi_2}-\Psi_2)\xi-\overline{\Psi^*_4}\rhoc,\label{eq:rell1}\\
&0=(\Psi_2+\overline{\Psi_2}-4\Psi_{11})\omega-({\Psi_2}-2\Psi_{11})\xi-\overline{\Psi^*_4}\rho+\Psi_{22}\varsigma+\overline{\Psi_4}\varsic,\label{eq:rell2}
\intertext{where \eqref{eq:bw-1sw1-a} and \eqref{eq:bw-1sw1-a} produce another algebraic relation}
&0=(\Psi_2-\overline{\Psi_2})\omega+(3\overline{\Psi_2}-2\Psi_{11})\zetac-(\Psi_2+\overline{\Psi_2}-4\Psi_{11})\xi
+\overline{\Psi^*_4}\rhoc+2\Psi_{22}\varsigma-\overline{\Psi^*_4}\frf.\label{eq:rell3}
\end{align}
On the other hand, compatibility of the $\tho'$-derivative of \eqref{eq:rhoeq}, the $\eth'$-derivative of \eqref{eq:taueq} with the Ricci equations \ric{6} and \ric{42}, producing
\be
\begin{aligned}
\ff\left[\frac{\tho'\w}{\w}\right.&\left.-\left(\frac{\w+3}{\w}\omega+\frac{\w+1}{\w}\zc\right)\varsic+2\muc-\frac{\w^2-1}{2\w}\cf\right]\\
&+\frac{\w-1}{\w}\left(\tho'\ff+2\Psi_{11}+\Lambda+\Dh\ef+\ef^2+\frac{(\w+1)^2}{2\w}\psi\psic\right)=0.\label{eq:masterric}
\end{aligned}
\ee
There is a qualitative difference in the treatment of the above equations for in the rank 1 vs.\ the rank 3 case, which we therefore treat separately.

\noindent{\em  Rank 1 case.} Given \eqref{rank1cond} the set \eqref{eq:bw-1sw0-a}-\eqref{eq:bw-1sw0-e} is found to be equivalent to $\varsigma\Psi_4^*$ being real, $\tho'\Psi_{11}=-2(\ff\Psi_{22}+\varsigma\Psi_4^*)$, and
\be\label{eq:muetc}
\mu=-\frac{3(\ff\Psi_{22}+\varsigma\Psi_4^*)}{4\Psi_{11}},\qquad 
\tho\Psi_{22}=-\ff\Psi_{22}-2\varsigma\Psi_4^*.
\ee
The set \eqref{eq:rell1}-\eqref{eq:rell3} gives
\be\label{eq:zetaetc}
\zc=\omega,\qquad \omega=\frac{2\varsigma\Psi_{22}-\ff\overline{\Psi_4^*}}{4\Psi_{11}},\qquad -3\varsigma\Psi_{22}+2\ff\overline{\Psi_4^*}+\varsic\overline{\Psi_4}=0,
\ee
such that \eqref{eq:masterric} becomes
\be\label{eq:mu-omega}
\ff\;\muc-3\varsic\omega=0.
\ee
Substituting $\mu$ from \eqref{eq:muetc} and $\omega$ from \eqref{eq:zetaetc}  into \eqref{eq:mu-omega} one gets
$\Psi_{22}(\ff^2+2\varsigma\varsic)=0$, whence $\Psi_{22}=0$. Then \eqref{eq:muetc} and \eqref{eq:mu-omega} imply $\mu=\varsigma\Psi_4^*=0$ and $\varsigma\omega=0$, such that $\zeta=\omega=\ff\Psi_{4}^*=0$ from the first two parts of \eqref{eq:zetaetc}. Since $(\varsigma,\frf)\neq (0,0)$ we obtain $\Psi_4^*=0$, which finishes the proof of statement B for this case. The last part of \eqref{eq:zetaetc} still gives
\be\label{eq:sigmaPsi4}
\varsigma\Psi_{4}=0.
\ee

\noindent{\bf Rank 3 case.} When \eqref{rank3cond} holds \eqref{eq:bw-1sw0-a}-\eqref{eq:bw-1sw0-e} can  be solved for $\mu$, $\tho'\Psi_{11}$, $\tho'\w$ and $\tho\Psi_{22}$, giving in particular
\beqn
\label{eq:mueq3}
&&\mu=\frac{2\w^2+1}{4\w}\frac{\ff\Psi_{22}}{\Psi_{11}}-\frac12\cf(\w-1),\\
\label{eq:tho'w}
&&\tho'\w=\frac{\w^2-1}{2}\cf-(\w^2-1)\frac{\ff\Psi_{22}}{2\Psi_{11}}.
\eeqn
Notice that the expression between brackets in the last term of \eqref{eq:masterric} is real 
and the complex conjugate of $(\w-1)/\w$ is $1-\w$. Hence, adding \eqref{eq:masterric} to $\w$ times its complex conjugate produces the relation
\be\label{eq:tho'wb}
\tho'\w=\frac{\w^2-1}{2}\cf-\frac{2\w(\w\muc+\mu)}{\w-1}.
\ee
Substitution of (\ref{eq:mueq3},\,\ref{eq:tho'w}) into \eqref{eq:tho'wb} yields $\ff\Psi_{22}(\w+1)(\w^2+\w+1)=0$, whence $\Psi_{22}=0$.\\
Next, \eqref{eq:biset1} can be solved for $\overline{\Psi_4^*}$, $\xi$ and $\zc$ in terms of $\omega$:
\be\label{eq:solveR4etc}
\overline{\Psi_4^*}=-\frac{4\Psi_{11}}{\w\ff}\omega,\qquad \xi=\frac{\w^2+\w+2}{(\w-1)\w}\omega,\qquad\zc=-\frac{\omega}{\w}.
\ee
By $\varsigma=0$  \ric{24} and \ric{42} respectively reduce to
\be\label{eq:d3upsee}
\eth\ef=\xi\ff+(\ff-\rhoc)\omega+\tau(\upsc+\ef),\qquad \eth\ups=-\xi\rho+\omega(\rho-\rhoc)-\psi(\ups-\upsc).
\ee
Substituting \eqref{eq:d3w} and \eqref{eq:d3upsee} into the $\eth$-derivative of \eqref{eq:ueq} and using (\ref{eq:rhoeq},\,\ref{eq:taueq},\,\ref{eq:ueq},\,\ref{eq:solveR4etc}) one obtains $\ff\omega(\w-1)=0$. Hence $\omega=0$, such that $\Psi_4^*=0$ and also $\xi=\zeta=0$ from  \eqref{eq:solveR4etc}.
\end{proof}

\noindent{\em B.2. $\Psi_4=0$}.

\begin{proof} In both rank 1 and 3 cases we already found $\zeta=\omega=0$. Hence \ric{5} and \ric{43} reduce to
\be\label{eq:ethtaupsi}
\eth\tau=\tau^2+\rho\overline{\lambda},\qquad \eth\psi=\psi^2+\ff\overline{\lambda}.
\ee
Substitution of \eqref{eq:d3w} and \eqref{eq:ethtaupsi} into the $\eth$-derivative of \eqref{eq:taueq} leads to $\ff\overline{\lambda}\W=0$. However, \bi{5}{3}{5}{3}{2} becomes $4\overline{\lambda}{\Psi_{11}}=\ff\Psi_4$, such that $\lambda=\ff\Psi_4=0$. Since we also found \eqref{eq:sigmaPsi4}
in the rank 1 case we obtain $\Psi_4=0$ in general, and theorem \ref{prop:rank 1-3} is proven.
\end{proof}

\end{document}